\documentclass[english,fleqn,10pt]{SelfArx}
\usepackage{lipsum}

\definecolor{color1}{RGB}{0,0,90} 
\definecolor{color2}{RGB}{0,20,20} 

\usepackage{hyperref} 
\hypersetup{hidelinks,colorlinks,breaklinks=true,urlcolor=color2,citecolor=color1,linkcolor=color1,bookmarksopen=false,pdftitle={Title},pdfauthor={Author}}

 \def\gsim{\lower.4ex\hbox{$\;\buildrel >\over{\scriptstyle\sim}\;$}}
 \def\lsim{\lower.4ex\hbox{$\;\buildrel <\over{\scriptstyle\sim}\;$}}
 \def\bl{\par\vskip 12pt\noindent}

 \def\aap{A\&A}
 \def\apj{ApJ}

 \def\basi{Bull. Astron. Soc. India}
 \def\ga{Geomagnetism and Aeronomy}
 \def\gafd{Geophys. Astrophys. Fluid Dyn.}
 \def\lrsp{Living Rev. Solar Phys.}
 \def\mnras{MNRAS}
 \def\nat{Nature}
 \def\prl{Phys. Rev. Lett.}

 \def\sp{Sol. Phys.}
 \def\aj{Astron. Rep.}
 \def\paj{Astron. Lett.}

\JournalInfo{Astronomy Letters, 2018, Vol. 44, No. 10} 
\Archive{}
\PaperTitle{
Solar cycle asymmetry as a consequence of fluctuations
in dynamo parameters
}
\Authors{L.~L.~Kitchatinov\textsuperscript* and A.~A.~Nepomnyashchikh}
\affiliation{\textit{Institute for Solar-Terrestrial Physics, Lermontov Str. 126A, Irkutsk, 664033, Russian Federation}}

\affiliation{*\textbf{E-mail}: kit@iszf.irk.ru}

\newcommand{\keywordname}{Keywords}

\Abstract{
The duration of activity growths in solar cycles is on average shorter than the duration of its declines. This asymmetry can result from fluctuations in dynamo parameters. A solar dynamo model with fluctuations in the $\alpha$-effect shows the statistical asymmetry which increases with both fluctuation amplitude and coherence time. An interpretation for the asymmetry origin is suggested, which predicts a correlation between the asymmetry measure and delay of the polar field reversals relative to the activity maxima. Data on the twelve latest solar cycles confirm such a correlation.
}

\begin{document}

\flushbottom 

\maketitle 


\thispagestyle{empty} 

\section{Introduction} 
The asymmetry of sunspot cycles is well known from observations (see, e.g., Hathawey et al. 1994; Obridko \& Nagovitsyn 2017; and references herein). Activity growth in individual cycles is usually less durable than its decline. Explanation of the asymmetry is a challenge to dynamo theory.

At first glance, it seems natural to relate the asymmetry to some nonlinear effect of the solar dynamo (Weiss et al. 1984). Field oscillations in linear (kinematic) dynamo models, as well as linear oscillations in general, are harmonic. Anharmonic undamped oscillations are typically not linear. Various nonlinearities are known in solar and stellar dynamos including magnetic modification of the differential rotation (Malkus \& Proctor 1975) and the meridional flow (Cameron \& Sch\"ussler 2012), suppression of field generation due to conservation of magnetic helicity (Kleeorin et al. 2003), modification of the   convective turbulence by the magnetic field (Kitchatinov et al. 1994), and others. Asymmetric magnetic cycles were indeed found in some models with sufficiently strong nonlinearities (e.g., Pipin \& Kosovichev 2011). However, the rotation rate of the Sun exceeds the threshold value for the onset of the global dynamo by about 10\% only (Metcalfe \& van\,Saders 2017; Kitchatinov \& Neponmyashchikh 2017a). The solar dynamo is, therefore, weakly nonlinear.

Recently, another possibility for the origin of solar cycles asymmetry has been  revealed: the asymmetry can result from fluctuations in dynamo parameters, namely, from fluctuations in the $\alpha$-parameter of the $\alpha$-effect. The fluctuations are caused by irregular variations in areas of the solar active regions and in angles of their tilts relative to the lines of latitude (Olemskoy et al. 2013). The consequences of the fluctuations depend substantially on the phase of an activity cycle in which they occur (Nagy et al. 2017). Therefore, it is not surprising that a dynamo model with fluctuating parameters shows asymmetry between the phases of rise and decline of the magnetic cycles (Kitchatinov et al. 2018). The asymmetry in the model, as well as in the observed solar cycles, is of a statistical nature: the rise phase is not shorter than the decline phase in all cycles, but the ratio of these phase durations is on average smaller than one.

This paper finds the dependence of the modelled asymmetry on the fluctuation para\-me\-ters and suggests its pictorial interpretation. Regularities shown by the model are compared with observations.
\section{Dynamo model}
\subsection{Outline of the model design}
Our dynamo model belongs to the so-called flux-transport models initiated by Durney (1995) and Choudhuri et al. (1995). This name reflects the significance of magnetic field advection by the meridional flow. Models of this type provide close agreement with solar observations (Jiang et al. 2013).

The numerical model permits computation of large-scale magnetic field dynamics in a spherical convective envelope. It differs from our earlier publications (Kitchatinov \& Nepomnyashchikh 2017a,b) only by allowance for the fluctuations in the $\alpha$-effect. Dynamo equations, boundary conditions, profiles of the angular velocity and the meridional flow and other model specifications have been discussed in detail in those papers. The way for allowance for the fluctuations has also been discussed earlier (Kitchatinov et al. 2018). Nevertheless, we repeat the discussion because the main subject with this paper -- solar cycle asymmetry -- is caused in the model exclusively by the fluctuations.

To account for the fluctuations in the $\alpha$-effect, the $\alpha$-para\-me\-ter in the poloidal field equation is changed as follows
\begin{equation}
    \alpha \rightarrow \alpha\left( 1 + \sigma s(t)\right),
    \label{1}
\end{equation}
where $\sigma$ is the relative amplitude of the fluctuations and $s(t)$ is a random function of time of the order one. Following Rempel (2005), the random process $s(t)$ is modelled by solving the system of $n$ ordinary differential equations,
\begin{eqnarray}
    \frac{\mathrm{d} s}{\mathrm{d} t} &=& -\frac{n}{\tau} \left( s - s_1\right),
    \nonumber \\
    \frac{\mathrm{d} s_1}{\mathrm{d} t} &=& -\frac{n}{\tau} \left( s_1 - s_2\right),
    \nonumber \\
    .\ .\ . &&
    \nonumber \\
     \frac{\mathrm{d} s_{n-1}}{\mathrm{d} t} &=& -\frac{n}{\tau}
     \left( s_{n-1} - \sqrt{\frac{2\tau}{\Delta t}}\ \hat{g}\right),
    \label{2}
\end{eqnarray}
in line with the dynamo equations. In these equations, $\tau$ is the correlation time, $\Delta t$ is the time-step in numerical integration of the equations, and $\hat{g}$ is a normally distributed random number with zero mean and {\sl rms} value equal to one. The value of $\hat{g}$ is renovated on each time step independently of its previous value. Solution of the Eqs.\,(\ref{2}) simulates a continuous function of time whose $n$-th order derivative is however discontinuous.    For $n = 1$ or $n = 2$ and under the condition of $\Delta t \ll \tau$, which is satisfied in all our computations, the correlation function for the random process of Eqs.\,(\ref{2}) can be derived analytically (Olemskoy \& Kitchatinov 2013). The coefficient in front of the random number $\hat{g}$ in the last of the equations (\ref{2}) is chosen so that the {\sl rms} value $\langle s^2\rangle^{1/2} = 1$ for the analytical correlations. $n = 3$ in the computations with this paper.

The $\alpha$-effect in our model is non-local. The poloidal field generation near the surface is related to the toroidal field near the base of the convection zone. The non-local formulation avoids the catastrophic quenching of the $\alpha$-effect caused by the conservation of magnetic helicity (Kitchatinov \& Olemskoy 2011). It also corresponds to the Babcock-Leighton mechanism for the poloidal field formation by the buoyant rise of toroidal fields from the base of the convection zone (see, e.g., Charbonneau 2010). The beginning of a magnetic cycle in our model is therefore defined as the instant of sign reversal of the toroidal field $B_\mathrm{t}$ near the base of the convection zone at 15$^\circ$ latitude where this field attains its largest strength in the course of its cyclic variations. Accordingly, the instant of a cycle maximum corresponds to the largest absolute value of $B_\mathrm{t}$ in the cycle.

The results to follow were computed with $\alpha = 0.174$\,m/s, which is 10\% above the threshold value for the onset of the dynamo in our model.
\subsection{How do fluctuations produce asymmetry?}
The magnetic fields of our model react to variations in the $\alpha$-parameter with a delay. The reaction of the poloidal (polar) field is delayed by about one year and four years more pass before the variation in $\alpha$ is felt by the near-bottom toroidal field $B_\mathrm{t}$ (Kitchatinov et al. 2018). The shape of a cycle is therefore controlled by the fluctuations in the growth phase of the cycle (fluctuations in the decline phase affect the amplitude - but not shape - of the next cycle).

Figure\,1 shows the magnetic field dependence on time in the computations with prescribed variations in $\alpha$ on the growth phase of a cycle. The variations were imposed between 3 and 4 years after the cycle onset. The computations were done for the reversed sign of $\alpha$ ($\alpha =-0.174$\,m/s) and for a change in the opposite direction by the same amount ($\alpha = 0.522$\,m/s) in this range of time. The cycle computed without variations in $\alpha$ is shown for comparison.

\begin{figure}[thb]
 \includegraphics[width=\columnwidth]{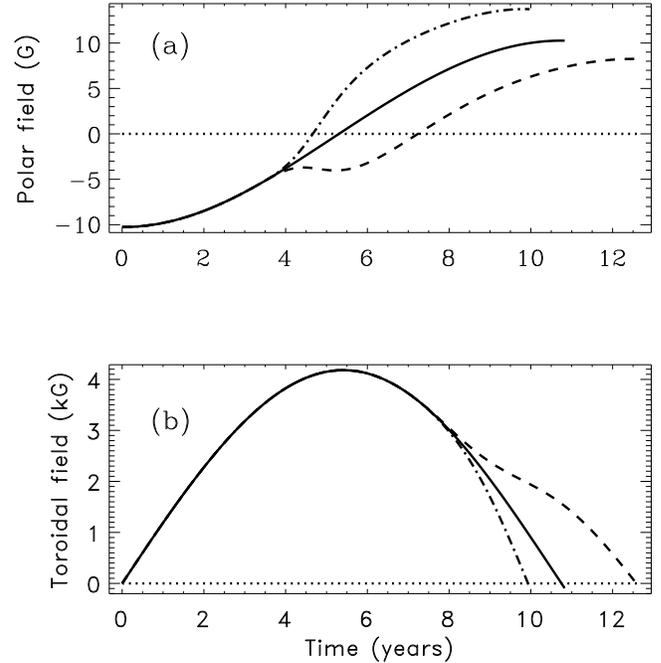}
 \caption{Field strength at the northern pole (a) and toroidal filed
     $B_\mathrm{t}$ at 15$^\circ$ latitude at the base of the convection zone (b) computed with  prescribed variations in $\alpha$ imposed in the range between 3 and 4 years after the cycle beginning. The dashed lines correspond to the reversal of the sign of $\alpha$ and the dashed-dotted lines -- to the change of $\alpha$ in the opposite direction by the same amount, i.e., to its threefold increase. For comparison, computation without variation in $\alpha$ is shown by the full line.
        }
    \label{f1}
\end{figure}

Without variations in $\alpha$, the computed cycle is symmetric: durations of the growth and decline phases are equal. The variations in $\alpha$ produce asymmetry. The decline phase is relatively short for positive (increase in $\alpha$) and relatively long for negative (decrease in $\alpha$) variations. As already mentioned, the toroidal field reaction delays by about 5 years relative to its causal variations. The growth branches are therefore identical in all three cases.

Figure 1 is easy to interpret in terms of the basic effects of the $\alpha\Omega$-dynamo. On the growth phase, the poloidal field opposite to the field of the previous activity minimum is produced by the $\alpha$-effect. This results in the poloidal field reversal near the activity maximum. Hereafter, the differential rotation produces a toroidal field opposite to the existing one and the magnetic activity declines. Positive fluctuation in $\alpha$ leads to an earlier reversal of the polar (poloidal) field so that the following decline of activity goes faster and the decline phase shortens. Negative fluctuation, on the contrary, lead to a delay in polar field reversal and extends the decline phase. Note that the positive fluctuation shortens the decline phase by a smaller amount compared to its prolongation by the negative fluctuation. Multiple positive and negative fluctuations, therefore, lead to a statistical mean asymmetry with relatively long epochs of activity decline.

This interpretation implies a correlation between the delay, $T_\mathrm{rev} - T_\mathrm{gr}$, of the reversal instant ($T_\mathrm{rev}$) relative to the activity maximum (coinciding with the duration $T_\mathrm{gr}$ of the growth phase) and the asymmetry parameter $T_\mathrm{gr}/T_\mathrm{dec}$ ($T_\mathrm{dec}$ is the decline phase duration). The correlation is indeed found in both the model computations and observational data.
\section{Results and discussion}
Mean durations of the growth and decline phases in magnetic cycles computed for various relative amplitudes and correlation times of $\alpha$-fluctuations are listed in Table\,1. Each averaging is performed over 4000 computed cycles. The statistics are representative. Their further extension does not change the results. The computations with various $\sigma$ and $\tau$ serve to ascertain that the modelled asymmetry is a regular reproducible phenomenon and that, as it should be, the asymmetry increases with either amplitude and mean duration of the fluctuations. Only for the smallest amplitude and correlation time of Table\,1, the sense of the modelled small asymmetry is opposite to observations (i.e. growth is longer than decline on average).

\bl {\small
\begin{description}
\item {\bf Table~1.}
Average durations of the growth and decline phases of dynamo-cycles computed with various amplitudes $\sigma$ and correlation times $\tau$ of the fluctuations in the $\alpha$-effect ($P_\mathrm{rot} = 25.4$\,days is the solar rotation period).
\end{description}
\centerline{
\begin{tabular}{c|c|c|c}
\hline
$\sigma$ & $\tau /P_\mathrm{rot}$ & $\langle T_\mathrm{gr}\rangle$\,(years) & $\langle T_\mathrm{dec}\rangle$\,(years) \\
\hline
         & 0.5 & 5.40 & 5.36 \\
\cline{2-4}
       1 & 1.0 & 5.39 & 5.39 \\
\cline{2-4}
         & 1.5 & 5.39 & 5.41 \\
\cline{2-4}
         & 2.0 & 5.37 & 5.45 \\
\hline
         & 0.5 & 5.37 & 5.45 \\
\cline{2-4}
       2 & 1.0 & 5.31 & 5.59 \\
\cline{2-4}
         & 1.5 & 5.21 & 5.70 \\
\cline{2-4}
         & 2.0 & 5.16 & 5.76 \\
\hline
         & 0.5 & 5.29 & 5.60 \\
\cline{2-4}
       3 & 1.0 & 5.14 & 5.80 \\
\cline{2-4}
         & 1.5 & 5.17 & 5.90 \\
\cline{2-4}
         & 2.0 & 5.22 & 5.94 \\
\hline
\end{tabular}}}
\bl

The Babcock-Leighton mechanism is related to observable properties of sunspots. Amplitude of fluctuations in the corresponding $\alpha$-effect can, therefore, be estimated from sunspot data. Estimations by Olemskoy et al. (2013) give the value of $\sigma \simeq 2.7$. The fluctuations mean duration $\tau$ can be inferred from a comparison of computed cycle periods with statistics of the observed solar cycle periods. Comparison with 36 periods of direct observations (https://www.ngdc. noaa.gov/stp/space-weather/solar-data/solar-indices/sunspot-\\ numbers/cycle-data/table\_cycle-dates\_maximum-minimum.\\txt)
and with 119 periods reconstructed by Nagovitsyn et al. (2015) from activity proxies give the correlation time $\tau \approx P_\mathrm{rot}$ (Kitchatinov et al. 2018). The discussion to follow concerns, therefore, the case of $\sigma = 3$ and  $\tau /P_\mathrm{rot} =1$ from Table\,1, which is closest to these estimations.

\begin{figure}[thb]
 \includegraphics[width=\columnwidth]{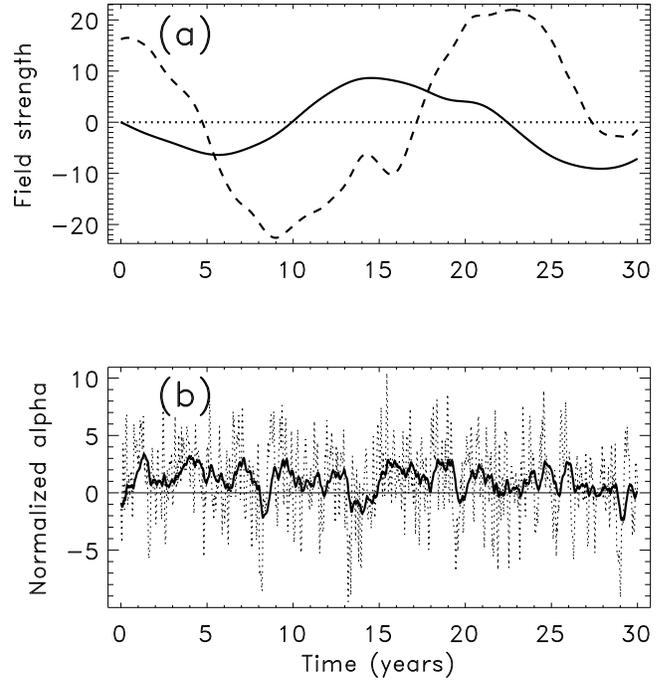}
 \caption{(a) Polar field (Gauss, dashed line) and the toroidal field
     $B_\mathrm{t}$ (kilo-Gauss, full line) for a computation fragment of the dynamo model with fluctuating $\alpha$-effect. \\ (b) Normalized $\alpha$-value $1 + \sigma s(t)$ of Eq.\,(\ref{1}) (dashed) and its annual running mean (full line). The cycle in the range between 10 and 23 years is highly asymmetric.
        }
    \label{f2}
\end{figure}

Figures 2 and 3 give an example of a strongly asymmetric cycle from our computations. Figure 2 agrees with the suggested explanation of the asymmetry and with Fig.\,1 of the preceding Section. The pronounced asymmetry of the cycle between 10 and 23 years in Fig.\,2 was in all probability caused by the negative fluctuation in $\alpha$ just before the cycle maximum. Due to the fluctuation, the reversal of the polar field, which would occur near the cycle maximum, was delayed. There was even a temporary increase in strength of the polar field before its reversal. As a result, the toroidal field decline was slow and the cycle turned out to be highly asymmetric.

\begin{figure}[h]
 \includegraphics[width=\columnwidth]{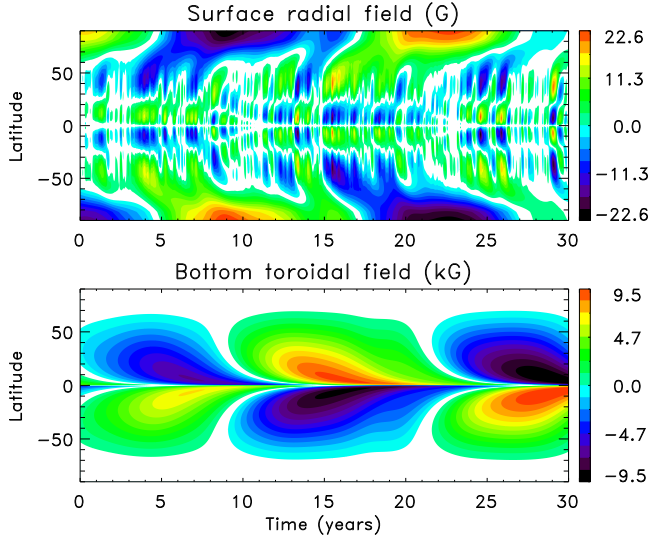}
 \caption{Time-latitude diagrams of the surface radial field (top panel) and
    the near-bottom toroidal field (bottom) for the same fragment as Fig.\,2.
        }
 \label{f3}
\end{figure}

The delay in polar reversal was associated with the surge of the poloidal field of  polarity of the previous minimum from low latitudes, where the $\alpha$-effect operates, to the poles (Fig.\,3). Such old-polarity surges from non-Joy active regions (correspond to negative $\alpha$-fluctuations in our model) are often met in observations (Jiang et al. 2015; Mordvinov et al. 2016).

The fine structure of the radial field at low latitudes seen in the upper panel of Fig.\,3 results from the fluctuations in $\alpha$. The fine structure is smoothed-out by turbulent diffusion as the field spreads to high latitudes and the polar field varies smoothly with time. Variation of the toroidal field in Figs.\,2 and 3 is smooth also. Nevertheless, false cycles in which $B_\mathrm{t}$ altered sign twice in a short time - sometimes shorter than one year - have been met in our computations. Such cycles usually come in pairs and they are distinguished  by the absence of polar field reversals. Obviously, the false cycles are caused by the fluctuations and they occur when after the change of sign of $B_\mathrm{t}$ it reverses again for a short time. About 2\% of computed cycles were false. The false cycles were excluded from the computational statistics and all the results to follow refer to such \lq cleaned' statistics (this practically did not affect the results, however).

\begin{figure}[htb]
 \includegraphics[width=\columnwidth]{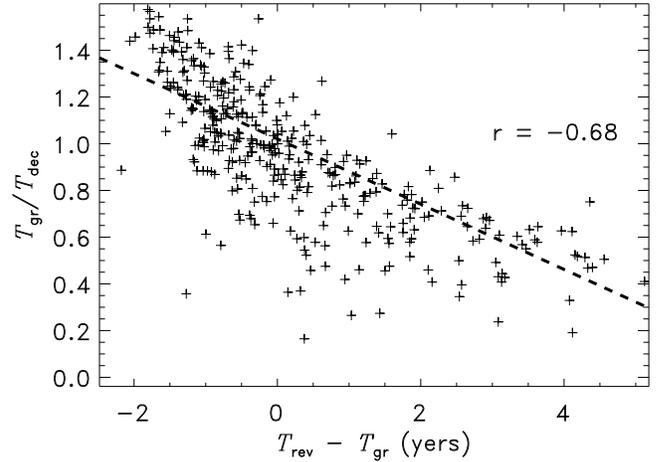}
 \caption{Anti-correlation between the asymmetry parameter
     $T_\mathrm{gr}/T_\mathrm{dec}$ and the delay of the instant $T_\mathrm{rev}$ of the polar field reversal relative to the time $T_\mathrm{gr}$ of the magnetic cycle maximum in the dynamo model. All times are measured from the instants of the cycles' onset. The correlation coefficient is $r = -0.68$. The dashed line shows the linear fit. Every tenth cycle only is shown by a cross to avoid jam in the Figure.
        }
    \label{f4}
\end{figure}

Our interpretation of the cycle asymmetry implies its relation to the delay in the polar field reversals relative to the cycle maxima. Figure\,4 shows that such a statistical relation is indeed present in our model.

\begin{figure}[hbt]
 \includegraphics[width=\columnwidth]{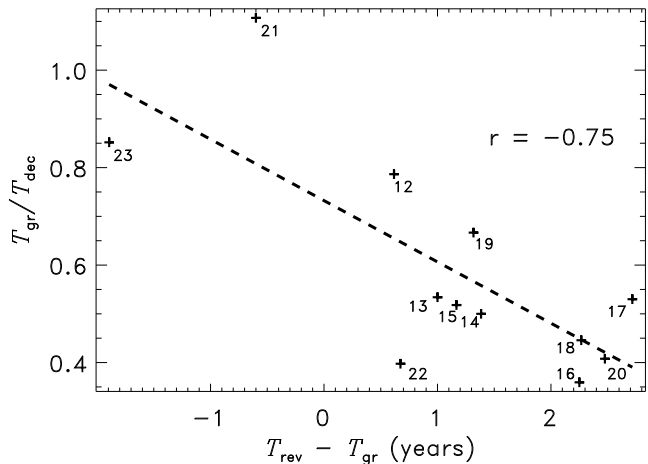}
 \caption{Anti-correlation between the asymmetry parameter and the
    delay of the polar field reversals relative to the maxima of solar cycles 12 to 23. The dashed line shows the best linear fit. The correlation coefficient is $r = -0.75$.
        }
    \label{f5}
\end{figure}

Solar observations offer the possibility for analysis of this relation for solar cycles 12 to 23. Dates of minima and maxima for these cycles were inferred from sunspot area data with exactly the same procedure as applied by Osipova \& Nagovitsyn (2017) (the same data were used and the same procedure of smoothing over 13 monthly  sunspot areas was applied). Dates of polar field reversals in the cycles 12 to 20 were taken from Makarov \& Sivaraman (1986) and for the cycles 20 to 23 - from data of the Wilcox Solar Observatory (http://wso.stanford.edu/Polar.html). Though the moderate statistics of 12 solar cycles does not allow a confident conclusion, it may be noticed that the observations-based Fig.\,5 shows the same tendency as the model computations of Fig.\,4.

Judging from Figs.\,1 and 2, a correlation between the cycle asymmetry and the growth-phase averaged value of $\alpha$ could also be expected. Weak correlation of this type is actually present in the model but its coefficient $r = 0.34$ is small. This is probably because the short-term positive and negative fluctuations in $\alpha$ may balance each other but their opposite in sense but different in value contributions to the asymmetry are not balanced (Fig.\,1).

A correlation between asymmetry and cycle period is not present in either the model computations or the observations.
\section{Concluding remarks}
Large fluctuations are inherent in the Babcock-Leighton mechanism for generation of the solar poloidal field. The amplitude of fluctuations in the corresponding dynamo parameter exceeds its mean value (Olemskoy et al. 2013). The reaction of the magnetic field to fluctuations in the mechanism of its generation can therefore be nonlinear: fluctuations in the $\alpha$-effect with zero mean produce unbalanced asymmetry in magnetic cycles (Fig.\,1). Therefore, fluctuations in the dynamo can be the reason - possibly not the sole one - for the observed asymmetry of solar cycles.

The suggested interpretation of the observed asymmetry leads to the anticipation of its correlation with the delay in polar field reversals relative to the activity maxima. Predictions of this type are infreuent in dynamo theory (see, however, Schatten et al. 1978; Choudhuri et al. 2007). The opinion has even been expressed  that the theory does not have predictive power (Tobias et al. 2006). Observational tests of the predictions are therefore important for the choice of an adequate model for the solar dynamo. Further observational tests may help to improve the model used in this paper.
\phantomsection
\section*{Acknowledgments}
This work was supported by budgetary funding of Basic Research program II.16 and by the Russian Foundation for Basic Research (project 16-02-00090).
\phantomsection
\section*{References}
\begin{description}
\item{} Cameron,~R.\,H., \& Sch\"ussler,~M. 2012, \aap\ {\bf 548}, 57
\item{} Charbonneau,~P. 2010, \lrsp\ {\bf 7}, 3
\item{} Choudhuri,~A.\,R., Sch\"ussler,~M., \& Dikpati,~M. 1995, \aap\ {\bf 303}, L29
\item{} Choudhuri,~A.\,R., Chatterjee,~P., \& Jiang,~J. 2007, \prl\ {\bf 98}, 131103
\item{} Durney,~B.\,R. 1995, \sp\ {\bf 160}, 213
\newpage
\item{} Hathawey,~D.\,H., Wilson,~R.\,M., \& Reichmann,~E.\,J. 1994, \sp\ {\bf 151}, 177
\item{} Jiang,~J., Cameron,~R.\,H., Schmitt,~D., \& Isik,~E. 2013, \aap\ {\bf 553}, A128
\item{} Jiang,~J., Cameron,~R.\,H., \& Sch\"ussler,~M. 2015, \apj\ {\bf 808}, L28
\item{} Kitchatinov,~L.\,L., \& Nepomnyashchikh,~A.\,A. 2017a,\\ \mnras\ {\bf 470}, 3124
\item{} Kitchatinov,~L.\,L., \& Nepomnyashchikh,~A.\,A. 2017b, \paj\ {\bf 43}, 332
\item{} Kitchatinov,~L.\,L., \& Olemskoy,~S.\,V. 2011, \paj\ {\bf 37}, 286
\item{} Kitchatinov,~L.\,L., R\"udiger,~G., \& K\"uker,~M. 1994, \aap\ {\bf 292}, 125
\item{} Kitchatinov,~L.\,L., Mordvinov,~A.\,V., \& Nepomnyashchikh,~A.\,A. 2018, \aap , in press, arXiv: 1804.02833 (2018).
\item{} Kleeorin,~N., Kuzanyan,~K., Moss,~D., Rogachevskii,~I., Sokoloff,~D., \& Zhang,~H. 2003, \aap\ {\bf 409}, 1097
\item{} Makarov,~V.\,I., \& Sivaraman,~K.\,R. 1986, \basi\ {\bf 14}, 163
\item{} Malkus,~W.\,V.\,R., \& Proctor,~M.\,R.\,E. 1975, J. Fluid. Mech. {\bf 67}, 417
\item{} Metcalfe,~T.\,S., \& van\,Saders,~J. 2017, \sp\ {\bf 292}, 126
\item{} Mordvinov,~A.\,V., Pevtsov,~A.\,A., Bertello,~L., \& Petri,~G.\,J.\,D. 2016, Solar-Terrestrial Physics {\bf 2}, 3
\item{} Nagovitsyn,~Yu.\,A., Georgieva,~K., Osipova,~A.\,A., \& Kuleshova,~A.\,I. 2015, \ga\ {\bf 55}, 1081
\item{} Nagy,~M., Lemerle,~A., Labonville,~F., Petrovay,~K., \& Charbonneau,~P. 2017, \sp\ {\bf 292}, 167
\item{} Obridko,~V.\,N., \& Nagovitsyn,~Yu.\,A. 2017, Solar activity, cyclicity and forecasting methods (St-Petersburg: BBM Press), p.169 (in Russian)
\item{} Olemskoy,~S.\,V., \& Kitchatinov,~L.\,L. 2013, \apj\ {\bf 777}, 71
\item{} Olemskoy,~S.\,V., Choudhuri,~A.\,R., \& Kitchatinov,~L.\,L. 2013, \aj\ {\bf 57}, 458
\item{} Osipova,~A.\,A., \& Nagovitsyn,~Yu.\,A. 2017, \ga\ {\bf 57}, 1092
\item{} Pipin,~V.\,V., \& Kosovichev,~A.\,G. 2011, \apj\ {\bf 741}, 1
\item{} Rempel,~M. 2005 \apj\ {\bf 631}, 1286
\item{} Schatten,~K.\,H., Scherrer,~P.\,H., Svalgaard,~L., \& Wilcox,~J.\,M. 1978, Geophys. Res. Lett. {\bf 5}, 411
\item{} Tobias,~S., Hughes,~D., \& Weiss,~N. 2006, \nat\ {\bf 442}, 26
\item{} Weiss,~N.\,O., Cattaneo,~F., \& Jones,~C.\,A. 1984, \gafd\ {\bf 30}, 305
\end{description}
\end{document}